\documentclass[letterpaper]{article}%
\usepackage{amssymb}
\usepackage{amsfonts}
\usepackage{amsmath}
\usepackage{hyperref}
\usepackage{graphicx}
\usepackage{float}%
\DeclareMathSymbol{\mrq}{\mathord}{operators}{`'}
\begin{document}

\title{Scattering Shadows}
\author{Thomas Curtright$^{\S }$ and Gaurav Verma$^{\flat}$\medskip\\Department of Physics, University of Miami, Coral Gables, FL 33124}
\date{}
\maketitle

\begin{abstract}
\textit{We discuss the regions forbidden to classical scattering trajectories
by repulsive potentials. \ We give explicit results for the asymptotic form of
these regions, far from the scattering center, in terms of the scattering
angle function.}

\end{abstract}

\section*{Introduction}

Consider particles of energy $E>0$ incident on a repulsive potential $V$ with
$E<V_{\max}$. \ The resulting classical scattering trajectories due to
smoothly varying, continuous potentials may exhibit \textquotedblleft
shadows\textquotedblright\ (classically forbidden regions) \cite{Refs}. \ The
structure of these shadows is discussed here, in a non-relativistic context
for various isotropic potentials, with particular attention to their
asymptotic form.

If $V\left(  r\right)  $ is falling monotonically but is non-vanishing for
\emph{all} distances from the fixed scattering center, the transverse size of
the scattering shadow may grow without bound as the downstream distance
increases in the forward direction, even if the potential becomes arbitrarily
weak at large distances. \ This is demonstrated here for various potentials.
\ In our opinion, this accounts for divergent total cross-sections in terms of
physical effects more easily visualized and more intuitively understood than,
say, angular integrations of differential cross-sections with singular small
angle behavior, or integrations over unbounded impact parameters. \ 

\section*{Classical Analysis}

All scattered particle trajectories are determined by the usual methods of
energy and angular momentum conservation, as exhibited here:%
\begin{equation}
E=\frac{m}{2}\left(  \frac{dr}{dt}\right)  ^{2}+\frac{L^{2}}{2mr^{2}}+V\left(
r\right)  \ ,\ \ \ L^{2}=m^{2}r^{2}\left(  \frac{d\vartheta}{dt}\right)  ^{2}%
\end{equation}
where $r$ is the distance of the particle from the fixed scattering center,
and the angle $\vartheta$\ is measured from the incoming direction of the
particle. \ (Please see the scattering diagram in Appendix A.) \ If the
particle's asymptotic speed is $v$ and it's impact parameter is $s$, then
\begin{equation}
E=mv^{2}/2\ ,\ \ \ L^{2}=m^{2}v^{2}s^{2}%
\end{equation}
If $a$ is the closest approach distance to the scattering center (i.e. $r$ at
periapsis) and $h=\left.  a\right\vert _{s=0}$ is the closest approach
distance for a head-on collision, with $h>0$ for $E<V_{\max}$, then
\begin{align}
E  &  =V\left(  h\right)  =\frac{L^{2}}{2ma^{2}}+V\left(  a\right)
\ ,\ \ \ L^{2}=2ms^{2}V\left(  h\right)  \ ,\ \ \ 1=\frac{s^{2}}{a^{2}}%
+\frac{V\left(  a\right)  }{V\left(  h\right)  }\\
\frac{m}{2}\left(  \frac{dr}{dt}\right)  ^{2}  &  =V\left(  h\right)  \left(
1-\frac{s^{2}}{r^{2}}\right)  -V\left(  r\right)  \ ,\ \ \ \frac{d\vartheta
}{dr}=\frac{d\vartheta/dt}{dr/dt}=\frac{s}{r^{2}}\frac{\left(  \mp1\right)
}{\sqrt{1-\frac{s^{2}}{r^{2}}-\frac{V\left(  r\right)  }{V\left(  h\right)  }%
}}%
\end{align}
with $\mp$ before/after periapsis and where $\vartheta$ is again measured from
the incident direction (say, the $-z$ axis). \ \smallskip

\noindent$\overline{^{\S }%
{\footnotesize curtright@miami.edu\ \ \ \ \ \ \ \ \ \ }^{\flat}%
{\footnotesize gxv211@miami.edu}}$\newpage

The scattering trajectories, as determined by $\vartheta\left(  r\right)  $,
therefore reduce to quadratures. \ Explicitly, upon changing variables to
$\sin\phi\equiv a/r$,%
\begin{align}
\vartheta\left(  r\right)   &  =\sqrt{1-\frac{V\left(  a\right)  }{V\left(
h\right)  }}\int_{0}^{\arcsin\left(  a/r\right)  }\frac{1}{f\left(
h,a,\phi\right)  }~d\phi\text{ \ for\ incoming }r\geq a\label{in}\\
&  \ \nonumber\\
\vartheta\left(  r\right)   &  =\sqrt{1-\frac{V\left(  a\right)  }{V\left(
h\right)  }}\int_{\arcsin\left(  a/r\right)  }^{\pi}\frac{1}{f\left(
h,a,\phi\right)  }~d\phi\text{\ \ for outgoing\ }r\geq a \label{out}%
\end{align}
where the denominator of the integrand is%
\begin{equation}
f\left(  h,a,\phi\right)  =\sqrt{1+\dfrac{V\left(  a\right)  \sin^{2}%
\phi-V\left(  a/\sin\phi\right)  }{V\left(  h\right)  \cos^{2}\phi}}
\label{integrand}%
\end{equation}
For repulsive potentials that fall monotonically to zero, $f$ is always real
and positive for all $\phi\in\left[  0,\pi\right]  $. \ The scattering angle
$\Theta$ as measured from the forward direction is given by $\Theta
=\pi-2\vartheta\left(  a\right)  $, hence%
\begin{equation}
\Theta=\pi-\sqrt{1-\frac{V\left(  a\right)  }{V\left(  h\right)  }}\int%
_{0}^{\pi}\frac{1}{f\left(  h,a,\phi\right)  }~d\phi\label{scatang}%
\end{equation}

Interesting examples are provided by inverse-power potentials, namely,%
\begin{equation}
V_{n}\left(  r\right)  =\frac{\kappa}{r^{n}}%
\end{equation}
with $\kappa>0$ and $n>0$, for which the post-periapsis angle is%
\begin{equation}
\vartheta_{n}\left(  r\right)  =\sqrt{1-\frac{h^{n}}{a^{n}}}\int%
_{\arcsin\left(  a/r\right)  }^{\pi}\frac{d\phi}{\sqrt{1+\dfrac{h^{n}}{a^{n}%
}\dfrac{\sin^{2}\phi-\sin^{n}\phi}{\cos^{2}\phi}}}\text{\ \ for outgoing\ }%
r\geq a
\end{equation}
In particular, for outgoing\ $r\geq a$, Coulomb potential results are%
\begin{align}
\vartheta_{1}\left(  r\right)   &  =2\arctan\left(  \sqrt{1-h/a}\right)
+2\arctan\left(  \frac{\sqrt{1-h/a}\sqrt{1-a/r}}{\sqrt{1+\left(  a-h\right)
/r}}\right) \\
\Theta_{1}  &  =\pi-2\vartheta_{1}\left(  a\right)  =\pi-4\arctan\left(
\sqrt{1-h/a}\right)
\end{align}
and inverse-square potential results are%
\begin{align}
\vartheta_{2}\left(  r\right)   &  =\sqrt{1-h^{2}/a^{2}}\left(  \pi
-\arcsin\left(  a/r\right)  \right) \\
& \nonumber\\
\Theta_{2}  &  =\pi-2\vartheta_{2}\left(  a\right)  =\pi\left(  1-\sqrt
{1-h^{2}/a^{2}}\right)
\end{align}
Some representative trajectories for these two particular cases are given in
the Figures of the following Section. \ Depending on the value of the impact
parameter, at specific points each trajectory is locally tangent to an
envelope curve, as is evident in the Figures.

By definition, for these examples or for any other repulsive potentials, at a
fixed energy $E<V_{\max}$ the \emph{scattering shadow} is the region
containing the scattering center and bounded by the trajectory envelope.
\ This is a region into which a \emph{classical} particle with energy $E$ is
forbidden to travel for \emph{all} impact parameters $s$ and corresponding
angular momenta $L$. \ Note that the length scales for this region are energy
dependent, as set by the value of $h$ for the potential in question. \ The
envelope can be obtained by a minimization procedure, in various ways, as
described in Appendix B.\newpage

\section*{Exemplary Trajectories}

First, consider $V_{1}=\kappa/r$. \ For this repulsive Coulomb scattering
example, the envelope of the trajectories is given exactly by \cite{1972,1991}%
\begin{equation}
\frac{z}{h}=-1+\frac{1}{4}\left(  \frac{y}{h}\right)  ^{2}\text{ \ \ where
\ \ }E=\frac{\kappa}{h}\label{CoulombShadowEqn}%
\end{equation}
The envelope is plotted along with representative trajectories in Figure 1. 

\begin{figure}[h]
\centering
\includegraphics[width=0.7\textwidth]{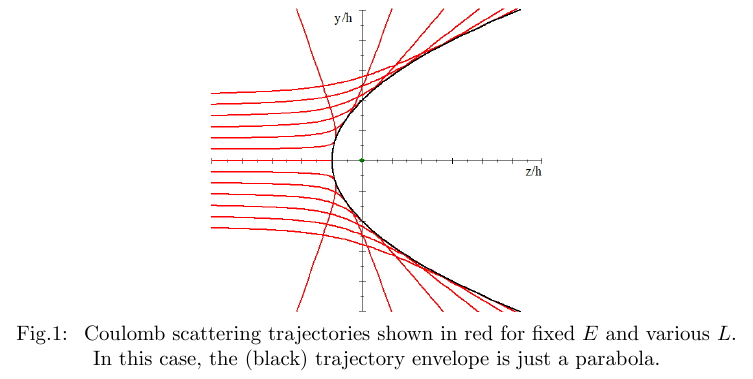}\end{figure}

Next, consider $V_{2}=\kappa/r^{2}$. \ For this inverse-square potential, in
Figure 2 the trajectories are only shown \emph{after} periapsis, and
the envelope curve is \emph{not} plotted. \ The envelope is not a parabola in
this case. 

\begin{figure}[h]
\centering
\includegraphics[width=0.85\textwidth]{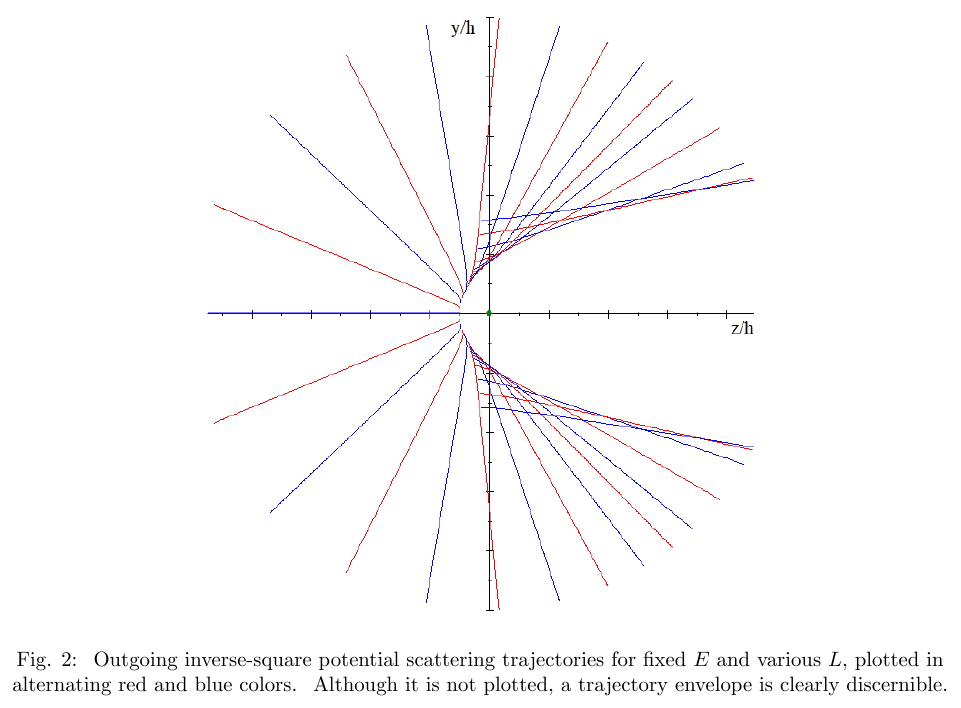}\end{figure}

\newpage

\section*{Analytic Shadow Procedures}

In principle, the shadow can be obtained analytically by the minimization
procedures described graphically in Appendix B. \ The third procedure
described therein is readily adapted to the Coulomb and inverse-square
potentials, given the explicit results for $\vartheta_{1}\left(  r\right)  $
and $\vartheta_{2}\left(  r\right)  $ as exhibited above. \ For a given
energy, hence a given $h$, and for a fixed $r$, the procedure is to minimize
$\vartheta$ by varying $s$, hence by varying $a$. \ 

For the Coulomb potential an elementary expression is obtained for the
critical $a$.%
\begin{equation}
\frac{\partial}{\partial a}\vartheta_{1}\left(  r\right)  =0\Longrightarrow
a=\frac{1}{2}h+\frac{1}{2}\sqrt{h\left(  4r-3h\right)  }
\label{CoulombCriticalA}%
\end{equation}
Evaluating $\vartheta_{1}\left(  r\right)  $ at this critical $a$ then gives
the enveloping curve solely as a function of $r$, for the chosen $h$. \ The
result is most easily expressed as $z=-r\cos\vartheta_{1}\left(  r\right)  $
written as a function of $y=r\sin\vartheta_{1}\left(  r\right)  $, as given in
(\ref{CoulombShadowEqn}) and shown as the parabola in the accompanying Figure 1.

For other inverse-power potentials the procedure is the same, but the final
result is not so simple. \ This is due to lack of a closed form expression for
the critical $a$ as a function of $r$, in contrast to the result in
(\ref{CoulombCriticalA}). The inverse-square potential illustrates this
difficulty. \ The condition to determine the envelope curve is%
\begin{equation}
\frac{\partial}{\partial a}\vartheta_{2}\left(  r\right)  =0\Longrightarrow
a\left(  a^{2}-h^{2}\right)  =h^{2}\sqrt{r^{2}-a^{2}}\left(  \pi-\arcsin
\frac{a}{r}\right)
\end{equation}
While the last equation is a fairly simple, closed-form, analytic result, the
solution for $a$ is not. \ Series solutions for $a\left(  r\right)  $ can be
obtained, say, by expanding about the $a=h$ head-on collision.\footnote{For
all repulsive potentials considered in this paper, near to $z=-h$ \ the
envelope is approximated by a parabolic function of $y$, similar to
(\ref{CoulombShadowEqn}).} \ But in general, for $a=O\left(  h\right)  $
numerical results are more useful. \ For example, with dimensionless variables
$\rho=r/h$ and $\alpha=a/h$, the envelope condition for $\rho=2$ is $\left.
\sqrt{\rho^{2}-\alpha^{2}}\left(  \pi-\arcsin\frac{\alpha}{\rho}\right)
=\left(  \alpha^{2}-1\right)  \alpha\right\vert _{\rho=2}$. \ The numerical
solution is $\alpha=1.6150$, and therefore $\left(  z/h,y/h\right)  =\left.
\left(  -\rho\cos\vartheta_{2}\left(  r\right)  ,\rho\sin\vartheta_{2}\left(
\rho\right)  \right)  \right\vert _{\rho=2,\alpha=1.615}=\left(
0.31479,1.9751\right)  $.

Figure 3 plots several numerical points (circled in red) for the
upper branch of the shadow boundary surrounding the fixed scattering center
located at the origin (green). 

\begin{figure}[h]
\centering
\includegraphics[width=0.75\textwidth]{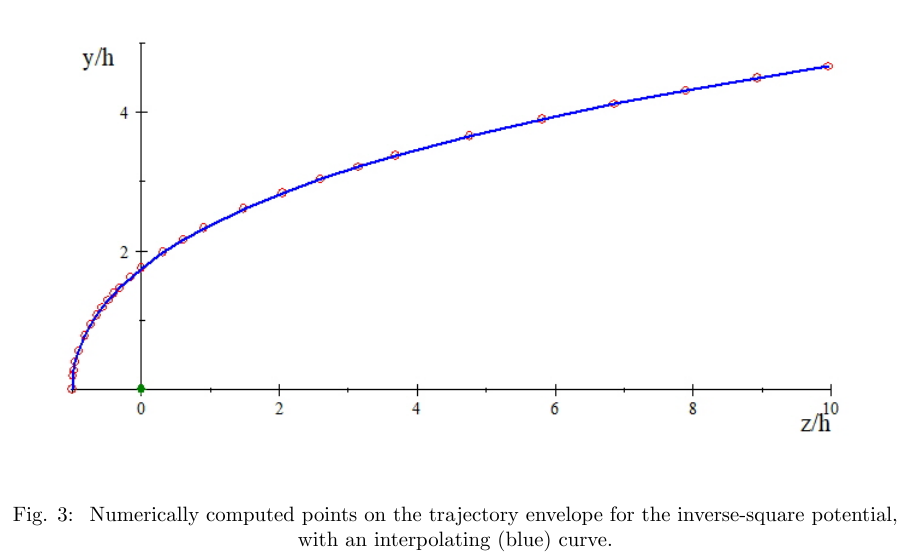}\end{figure}

However, in lieu of numerical analysis, far from the scattering center in the
downstream direction ($+z$ in the Figure) the asymptotic form of the envelope
can often be obtained as a simple closed-form expression.

\newpage

\section*{Exemplary Asymptotic Behavior}

The asymptotic form of the envelope, for $z\gg h$, follows from the behavior
of trajectories for small scattering angles, which are in turn the result of
large impact parameters. \ For $V_{n}\left(  r\right)  =\kappa/r^{n}$, with
$\kappa>0$ and $n>0$, the small scattering angle behavior is
\begin{equation}
\Theta\left(  s\right)  \underset{s\gg h}{\sim}\frac{h^{n}}{s^{n}}~b_{n}%
\ll1\ ,\ \ \ b_{n}=\sqrt{\pi}~\Gamma\left(  \tfrac{n+1}{2}\right)
/\Gamma\left(  \tfrac{n}{2}\right)
\end{equation}
as shown in Appendix C. \ For incident particles with \emph{any} impact
parameters striking such continuous repulsive potentials that fall off
smoothly with distance, after passing the point of periapsis the trajectories
eventually approach straight lines. \ Moreover, for \emph{large} impact
parameters these eventual straight lines are given simply by
\begin{equation}
y\underset{z\gg s\gg h}{\sim}s+z\tan\Theta\left(  s\right)  \sim
s+z~\Theta\left(  s\right)
\end{equation}
where $z$ is the downstream coordinate along the incident beam axis, $y$ is
the coordinate transverse to that axis, and $\Theta$ is the scattering angle.
\ In the first step exhibiting the asymptotic behavior the intercept of the
trajectory has been approximated by just $s$, while in the second step the
smallness of the scattering angle has also been used, both steps being good
approximations for large impact parameters.

One way to obtain the envelope for the trajectories is to minimize $y^{2}$ at
a fixed $z$ by varying $s$. \ The asymptotic behavior of the envelope then
follows from the condition $0=\partial y/\partial s\sim1+z~\Theta^{\prime
}\left(  s\right)  $. \ Thus, for the $V_{n}$ potential,
\begin{equation}
z\sim-\frac{1}{\Theta^{\prime}}=\frac{s}{n\Theta}=\frac{s^{n+1}}{nh^{n}b_{n}%
}\ ,\ \ \ s\sim\left(  nh^{n}b_{n}z\right)  ^{1/\left(  n+1\right)  }%
\end{equation}
The asymptotic form of the envelope is then obtained by evaluating $y$ subject
to this last relation between $s$ and $z$. \ The result is%
\begin{equation}
\frac{y_{\text{env}}}{h}\underset{z\gg s\gg h}{\sim}\frac{\left(  n+1\right)
s}{nh}\sim c_{n}~\left(  \frac{z}{h}\right)  ^{1/\left(  n+1\right)  }%
\end{equation}
for the $V_{n}$ potential, where\footnote{Compare this result to the area of a
unit sphere in $N$ spatial dimensions: $\ \Omega_{N}=\frac{2\pi^{N/2}}%
{\Gamma\left(  N/2\right)  }$ so $c_{n}=\frac{n+1}{n}\left(  n\pi~\frac
{\Omega_{n}}{\Omega_{n+1}}\right)  ^{1/\left(  n+1\right)  }$.}%
\begin{equation}
c_{n}=\frac{n+1}{n}\left(  nb_{n}\right)  ^{1/\left(  n+1\right)  }=\frac
{n+1}{n}\left(  \frac{n\sqrt{\pi}~\Gamma\left(  \frac{n+1}{2}\right)  }%
{\Gamma\left(  \frac{n}{2}\right)  }\right)  ^{1/\left(  n+1\right)  }%
\end{equation}
For example: $c_{1}=2\ ,\ \ c_{2}=\frac{3}{2}\pi^{1/3}=2.196\,89\ ,\ \ c_{3}%
=\frac{4}{3}6^{1/4}=2.086\,78\ ,\ \ c_{4}=\frac{5}{4}\left(  3\pi\right)
^{1/5}=1.957\,78\ ,\ $and$\ c_{\infty}=1$. \ The maximum $c_{n}$ occurs at
$n=1.797\,296\,395$, for which $\left.  c_{n}\right\vert _{n=1.797\,296\,395}%
=2.202\,842\,270$. \ Therefore, for any finite $n>0$ the transverse size of
the shadow grows without bound as $z$ increases.

In particular, for the inverse-square potential,
\begin{equation}
y_{\text{env}}/h\underset{z\gg h}{\sim}\frac{3}{2}\left(  \pi z/h\right)
^{1/3} \label{V2Env}%
\end{equation}
This approximation to the envelope curve is significantly below the actual
envelope for $z=O\left(  h\right)  $, as should be expected since it vanishes
at $z=0$. \ But beyond that, this leading asymptotic term is also too high by
a small amount for large $z/h$, as is evident in Figure 4.
\ Nevertheless, the difference gives a relative error that is negligible for
very large $z/h$, and that goes to zero as $z/h\rightarrow\infty$. \ In fact,
the first correction to the leading asymptotic behavior of the envelope for
inverse-square scattering\footnote{Derivation of this correction is left as an
exercise for the interested reader.} is given by $-\frac{3}{8}\frac{1}{\left(
\pi z/h\right)  ^{1/3}}$. \ Or, upon incorporating this into a simple Pad\'{e}
approximant, an improved asymptotic approximation is
\begin{equation}
y_{\text{env}}/h\underset{z\gg h}{\sim}\frac{\frac{3}{2}\left(  \pi
z/h\right)  ^{1/3}}{1+\frac{1}{4}\left(  h/\pi z\right)  ^{2/3}}
\label{V2Pade}%
\end{equation}
Again as is evident in Figure 4, for $z/h>10$ or so, this gives a better fit
to the numerically exact curve. 

\begin{figure}[h]
\centering
\includegraphics[width=0.9\textwidth]{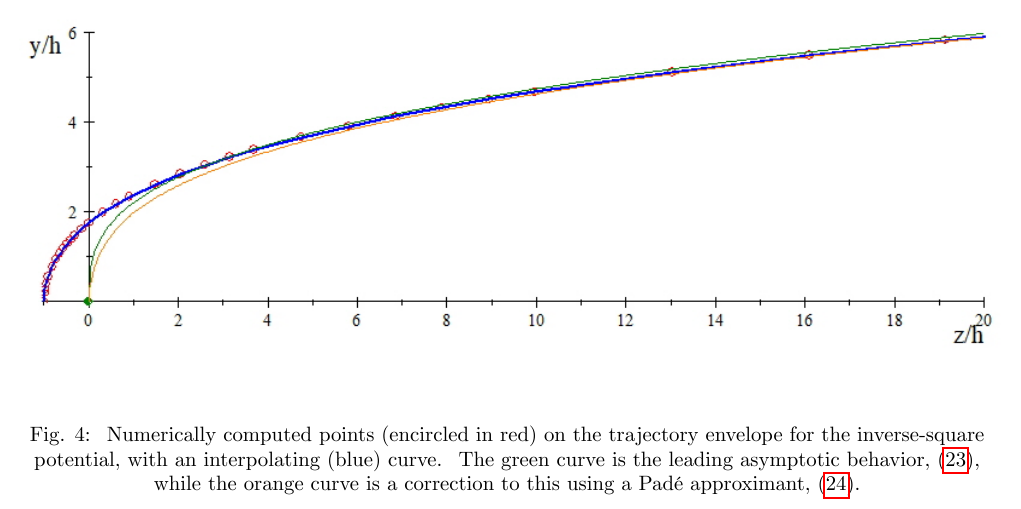}
\end{figure}

As another example, consider a Yukawa potential.%
\begin{equation}
V\left(  r\right)  =\frac{\kappa}{r}\exp\left(  -r/\lambda\right)
\end{equation}
In this case the relevant asymptotic behavior is given by the following, again
as taken from Appendix C.%
\begin{equation}
\Theta\underset{a\approx s\gg h,\lambda}{\sim}\sqrt{\frac{8\pi}{s\lambda}%
}h\exp\left(  -\frac{s}{\lambda}\right)  \ ,\ \ \ \frac{d}{ds}~\Theta
\underset{a\approx s\gg h,\lambda}{\sim}-\frac{1}{\lambda}~\Theta
\end{equation}%
\begin{equation}
z\sim-\frac{1}{\Theta^{\prime}}\sim\frac{\lambda}{\Theta}\sim\frac{\lambda}%
{h}\sqrt{\frac{s\lambda}{8\pi}}\exp\left(  \frac{s}{\lambda}\right)
\ ,\ \ \ s\sim\lambda\ln\left(  z/\lambda\right)
\end{equation}
The large $z$ form of the scattering envelope follows by the same methods as
used for the inverse-power potentials.%
\begin{equation}
y_{\text{env}}\underset{z\gg a\approx s\gg h,\lambda}{\sim}s+z~\Theta\left(
s\right)  \sim\lambda\ln\left(  z/\lambda\right)  +h\sqrt{\frac{8\pi}%
{\ln\left(  z/\lambda\right)  }}\sim\lambda\ln\left(  z/\lambda\right)
+O\left(  \frac{1}{\sqrt{\ln\left(  z/\lambda\right)  }}\right)
\end{equation}
For fixed $\lambda>0$ the transverse size of the shadow still grows without
bound as $z$ increases, but only as $\ln z$, hence more slowly than for any
inverse-power potential $V_{n}$ with $n>0$.

\section*{Total Cross-sections}

It is well-known that the \textquotedblleft total\textquotedblright%
\ cross-section $\sigma=\int\left(  \frac{d\sigma}{d\Omega}\right)  d\Omega$
is \emph{infinite} for the Coulomb potential when computed classically. \ It
is perhaps less well-known that this is true for \emph{all} inverse-power
potentials $V_{n}$ with $n>0$. \ Moreover, it is perhaps even less well-known
that $\sigma$ is also infinite for the Yukawa potential when computed
\emph{classically}. \ There are at least three ways to obtain these divergent
total cross-sections. \ These involve: \ Impact parameter integrals;
Integrations over singular small angle behavior; Scattering shadows. \ The
amount of insight and intuition provided by these three approaches differs
significantly, increasing with the listed ordering, in our opinion.

The first way to understand infinite $\sigma$ is to compute the total
cross-section in terms of the impact parameter, e.g. $\sigma=2\pi\int sds$ in
three spatial dimensions, and to use the fact that the upper limit of the
integral is $\infty$ if the potential is non-zero, albeit arbitrarily small,
for all finite distances no matter how far from the fixed scattering center.
However, this computation reveals next to nothing about the detailed structure
of the potential.\footnote{For anyone seeking to find an elementary physics
example that uses category theory ... well ... this is an \textquotedblleft%
\href{https://en.wikipedia.org/wiki/Abstract_nonsense}{abstract nonsense}%
\textquotedblright\ proof\ if we ever saw one.}

A more informative way to understand infinite $\sigma$ is to consider the
small angle behavior of $d\sigma/d\Omega$. \ If this differential
cross-section is sufficiently singular as $\Theta\rightarrow0$ then again
$\sigma$ is infinite due to the divergence of the integral $\int_{0}^{\pi
}\left(  d\sigma/d\Theta\right)  d\Theta$. \ This is the case for all the
aforementioned repulsive inverse-power potentials and also for the Yukawa
potential, when computed classically, as can be seen using the small angle
scattering results in Appendix C. \ This method encodes more information about
the potential than the impact parameter calculation, since it requires precise
knowledge about scattering behavior at small angles. \ But, in our opinion,
this computation is not easily visualized and does not impart much physical
intuition. \ Moreover, this approach relies on an infinite limit, namely, it
is necessary that the impact parameter $s\rightarrow\infty$ to produce
$\Theta\rightarrow0$ for the repulsive potentials under consideration.

On the other hand, the scattering shadow is a very intuitive way to understand
the physics that underlies $\sigma$. \ Again in our opinion, this approach
accounts for divergent total cross-sections in terms of effects very easily
visualized, namely, the unbounded growth of the shadow's transverse area as
the downstream distance from the scattering center increases. \ In addition,
this third way of thinking provides a clear picture of classical physics
effects even when observations are made at \emph{finite} downstream distances
since the shadow region is well-defined for all $z>-h$. \ Thus, once in hand,
the scattering shadow admits greater physical transparency, visualization, and
intuition than provided by either the impact parameter or scattering angle
considerations alone.

\section*{Summary}

The structure of scattering shadows\ (classically forbidden regions) was
discussed here, in a non-relativistic context for various isotropic
potentials, with particular attention to their asymptotic form. For the
monotonically falling, repulsive potentials considered, the transverse size of
the scattering shadow was shown to grow without bound as the downstream
distance increased in the forward direction, even though the potentials became
arbitrarily weak at large distances. This shadow growth accounted for
divergent total cross-sections in terms of physical effects more easily
visualized and more intuitively understood than either angular or impact
parameter integrations.

That said, as anyone familiar with light waves or quantum mechanics already
knows, these shadows do \emph{not} persist when point-particle classical
mechanics is set aside for more accurate descriptions of scattering processes.
\ Indeed, penetration of the classical shadow by diffracted light to form
\href{https://en.wikipedia.org/wiki/Arago_spot}{Poisson's spot} in the forward
direction was the experimentum crucis that distinguished electromagnetic waves
from Newton's particle theory of light \cite{Poisson spot}. \ Similar
phenomena have been observed for molecular beams \cite{More spot}, thereby
providing further evidence in favor of quantum mechanics. \ 

Perhaps the situation was best described in another context:

\begin{quote}
\textquotedblleft But in the end it's only a passing thing, this shadow
...\textquotedblright\ ---
\href{https://en.wikipedia.org/wiki/Samwise_Gamgee}{Sam},
in\textit{\ \href{https://en.wikipedia.org/wiki/The_Lord_of_the_Rings_(film_series)}{\textit{The
Lord of the Rings}}.}
\end{quote}

\noindent Nevertheless, to make a clear distinction between such alternative
scattering theories, a full appreciation of classical mechanics is required.
\ Hopefully the discussion given here has increased that appreciation.

\section*{Acknowledgements}

Discussions with Tom Kephart were greatly appreciated. \ This work was
supported in part by the United States Social Security Administration.

\newpage

\section*{Appendix A: \ Scattering Diagram}

Variables used to describe the classical scattering process are illustrated in
Figure 5 for a particular trajectory, as shown in blue.

\begin{figure}[h]
\centering
\includegraphics[width=0.62\textwidth]{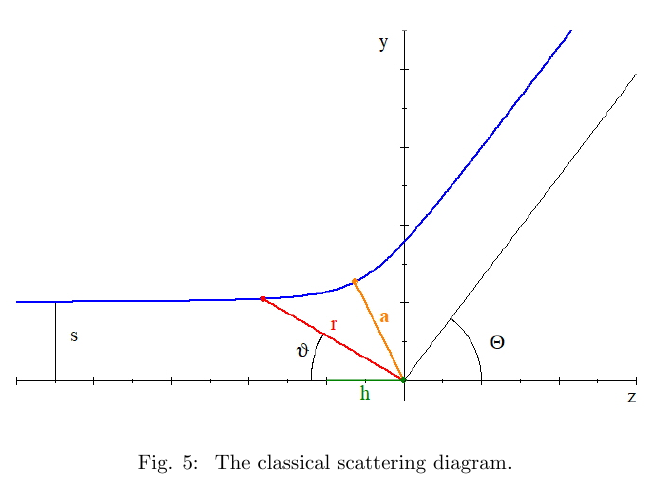}
\end{figure}

The coordinate frame has been chosen so that the trajectory lies in the
$yz$-plane, and the particle with energy $E$ is incident from infinity with
initial asymptotic velocity $\overrightarrow{v}=\sqrt{2mE}~\widehat{z}$. \ For
the trajectory shown, the impact parameter is $s$, the periapsis is $a$, and
the scattering angle is $\Theta$. \ The closest approach distance for a
head-on collision with the same energy $E$ is shown in the diagram as $h$.

\section*{Appendix B: \ Minimization Procedures}

For a given energy, the envelope can be obtained by varying the impact
parameter $s$ (hence varying $L=\sqrt{2mE}s$) to find the minimum transverse
distance $y$ at a fixed value of $z$, as shown graphically by trajectories
intersecting the vertical green lines in Figure 6.

\begin{figure}[h]
\centering
\includegraphics[width=0.62\textwidth]{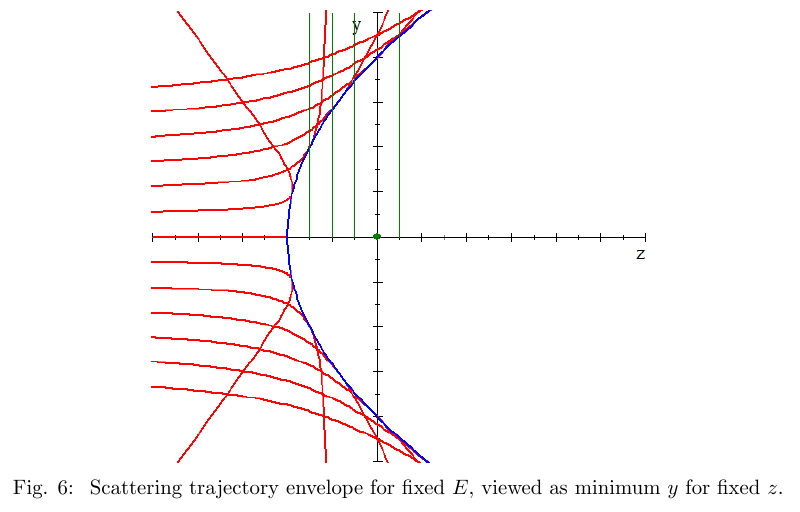}\end{figure}

Another, equivalent procedure to find the envelope is by varying the impact parameter
$s$ to find the minimum radial distance $r$ at a fixed value of the polar
angle $\theta$, as shown graphically by trajectories intersecting the green
rays in Figure 7.
 
\begin{figure}[h]
\centering
\includegraphics[width=0.62\textwidth]{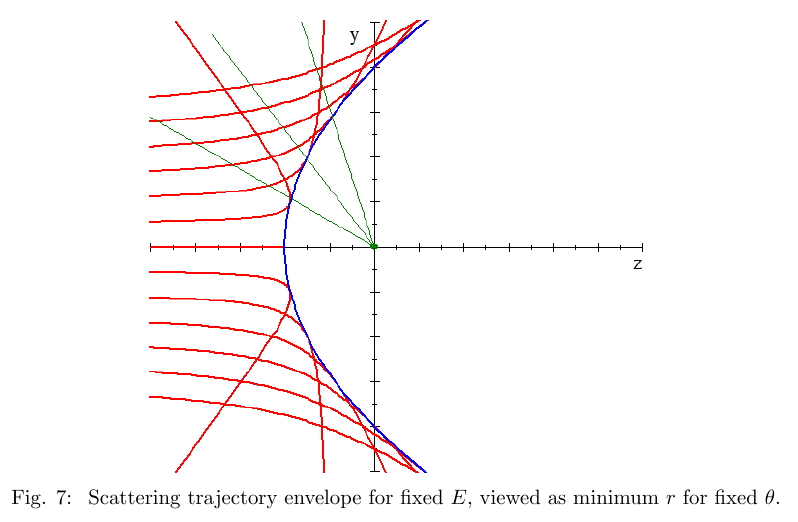}\end{figure}

Yet another procedure to find the envelope is by varying the impact parameter $s$ to find
the minimum polar angle $\theta$ at a fixed value of $r$, as shown graphically
by trajectories intersecting the green circles in the following Figure 8.

\begin{figure}[h]
\centering
\includegraphics[width=0.62\textwidth]{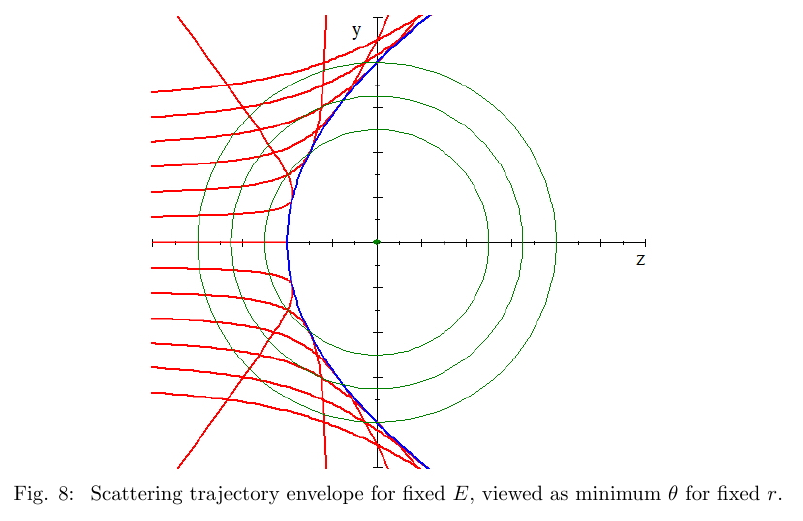}\end{figure}

\section*{Appendix C: \ Small Angle Scattering}

For large impact parameters $s\gg h$, hence periapsis distances $a\gg h$ and
therefore small $\Theta$, the scattering angle integral (\ref{scatang}) may be
approximated by expanding the integrand to first nontrivial order in $V\left(
a\right)  /V\left(  h\right)  $. \ Thus
\begin{equation}
\Theta\underset{a\gg h}{=}\pi-\sqrt{1-\frac{V\left(  a\right)  }{V\left(
h\right)  }}\int_{0}^{\pi}\left(  1-\dfrac{V\left(  a\right)  \sin^{2}%
\phi-V\left(  a/\sin\phi\right)  }{2V\left(  h\right)  \cos^{2}\phi}+O\left(
\frac{V\left(  a\right)  ^{2}}{V\left(  h\right)  ^{2}}\right)  \right)
d\phi\tag{C1}%
\end{equation}
Also expanding the $\sqrt{\cdots}$ prefactor gives%
\begin{equation}
\Theta\underset{a\gg h}{=}\frac{\pi}{2}\frac{V\left(  a\right)  }{V\left(
h\right)  }+\frac{1}{2}\int_{0}^{\pi}\dfrac{V\left(  a\right)  \sin^{2}%
\phi-V\left(  a/\sin\phi\right)  }{V\left(  h\right)  \cos^{2}\phi}%
d\phi+O\left(  \frac{V\left(  a\right)  ^{2}}{V\left(  h\right)  ^{2}}\right)
\tag{C2}%
\end{equation}
For inverse power potentials the remaining integrand is%
\begin{equation}
\dfrac{V_{n}\left(  a\right)  \sin^{2}\phi-V_{n}\left(  a/\sin\phi\right)
}{V_{n}\left(  h\right)  \cos^{2}\phi}=\frac{h^{n}}{a^{n}}\frac{\sin^{2}%
\phi-\sin^{n}\phi}{\cos^{2}\phi} \tag{C3}%
\end{equation}
which leads to an $n$-dependent integral without any further dependence on $h
$ or $a$.%
\begin{equation}
\Theta\underset{a\gg h}{=}\frac{h^{n}}{a^{n}}\left(  \frac{\pi}{2}+\frac{1}%
{2}\int_{0}^{\pi}\frac{\sin^{2}\phi-\sin^{n}\phi}{\cos^{2}\phi}d\phi\right)
+O\left(  \frac{h^{2n}}{a^{2n}}\right)  \tag{C4}%
\end{equation}
Note the integrand vanishes when $n=2$. \ Otherwise, direct evaluation of this
last integral gives\footnote{Therein lies a tale. \ For even $n=2k$, a
generating function for such integrals is given by
\[
\frac{2}{\pi x}\sum_{k=1}^{\infty}x^{k}\left(  \frac{\sqrt{\pi}~\Gamma\left(
\frac{2k+1}{2}\right)  }{\Gamma\left(  k\right)  }-\frac{\pi}{2}\right)
=\frac{1}{\left(  1-x\right)  ^{3/2}}-\frac{1}{1-x}%
\]
See \href{https://oeis.org/A173384}{https://oeis.org/A173384}.}%
\begin{equation}
\frac{1}{2}\int_{0}^{\pi}\frac{\sin^{2}\theta-\sin^{n}\theta}{\cos^{2}\theta
}\ d\theta=\frac{\sqrt{\pi}~\Gamma\left(  \frac{n+1}{2}\right)  }%
{\Gamma\left(  \frac{n}{2}\right)  }-\frac{\pi}{2} \tag{C5}%
\end{equation}
and hence the power-law potential result given in the text:%
\begin{equation}
\Theta\underset{a\gg h}{=}\frac{h^{n}}{a^{n}}\frac{\sqrt{\pi}~\Gamma\left(
\frac{n+1}{2}\right)  }{\Gamma\left(  \frac{n}{2}\right)  }+O\left(
\frac{h^{2n}}{a^{2n}}\right)  \tag{C6}%
\end{equation}

By comparison, the small scattering angle approximation for a Yukawa potential
leads to an integral that is more challenging. \ Let $V\left(  r\right)
=\frac{\kappa}{r}\exp\left(  -r/\lambda\right)  $ to find
\begin{equation}
\Theta\underset{a\gg h}{=}\frac{h}{2a}\exp\left(  \frac{h-a}{\lambda}\right)
\left(  \pi+\int_{0}^{\pi}\frac{\sin\phi}{\cos^{2}\phi}\left(  \sin\phi
-\exp\left(  \frac{a}{\lambda}\left(  1-\frac{1}{\sin\phi}\right)  \right)
\right)  d\phi\right)  +O\left(  \frac{V\left(  a\right)  ^{2}}{V\left(
h\right)  ^{2}}\right)  \tag{C7}%
\end{equation}
and eventually obtain, in Maple$^{\textregistered}$ notation,%
\begin{equation}
\Theta\underset{a\gg h}{=}\frac{h}{a}\exp\left(  \frac{h-a}{\lambda}\right)
\left(  -3+\sqrt{8\pi a/\lambda}\operatorname{hypergeom}\left(  \left[
-\frac{1}{2},\frac{3}{2}\right]  ,\left[  {}\right]  ,-\frac{\lambda}%
{2a}\right)  \right)  +O\left(  \frac{V\left(  a\right)  ^{2}}{V\left(
h\right)  ^{2}}\right)  \tag{C8}%
\end{equation}
For fixed $\lambda$, small angle scattering is again obtained for $a\approx
s\gg h$ or $\lambda$. \ But in this limit the hypergeometric function
simplifies to $\lim_{\lambda/a\rightarrow0}\operatorname{hypergeom}\left(
\left[  -\frac{1}{2},\frac{3}{2}\right]  ,\left[  {}\right]  ,-\frac{\lambda
}{2a}\right)  =1$. \ Thus%
\begin{gather}
\Theta\underset{a\approx s\gg h,\lambda}{\sim}\sqrt{\frac{8\pi}{s\lambda}%
}~h\exp\left(  -\frac{s}{\lambda}\right)  \ ,\ \ \ \frac{d}{ds}~\Theta
\underset{a\approx s\gg h,\lambda}{\sim}-\frac{1}{\lambda}\sqrt{\frac{8\pi
}{s\lambda}}~h\exp\left(  -\frac{s}{\lambda}\right) \tag{C9}\\
\frac{s}{\lambda}\underset{\Theta\ll1}{\sim}\frac{8\pi h^{2}}{\Theta
^{2}\lambda^{2}}\exp\left(  -\operatorname{LambertW}\left(  \frac{16\pi h^{2}%
}{\Theta^{2}\lambda^{2}}\right)  \right)  \tag{C10}%
\end{gather}
as used in the text. $\ $Also
\href{https://en.wikipedia.org/wiki/Lambert_W_function#Asymptotic_expansions}{recall
}$\operatorname{LambertW}\left(  w\right)  \underset{w\gg1}{\sim}\ln w-\ln\ln
w+o\left(  1\right)  $, so $w\exp\left(  -\operatorname{LambertW}\left(
w\right)  \right)  \underset{w\gg1}{\sim}O\left(  \ln w\right)  $.

\end{document}